\documentstyle[editedvolume]{jd} 
\begin{opening}
\title{INFRARED OBSERVATIONS OF GALAXY CLUSTERS}

\author{D. ELBAZ}
\institute{CEA - Service d'Astrophysique \\
Orme des Merisiers, 91191 Gif-sur-Yvette Cedex - France}

\end{opening}

\runningtitle{INFRARED OBSERVATIONS OF GALAXY CLUSTERS}

\begin{document}


\section{Introduction} 
The evolution of galaxy clusters from their formation due to the
merging of sub structures, the bulk of star formation and subsequent
chemical enrichment of the intra-cluster medium, is expected to be
quite recent (z$<$1-2) in the hierarchical clustering scenario (White
$\&$ Frenk 1991). 

Some evidence for enhanced star formation at z $\simeq$ 0.3-0.5 has
indeed been collected during the past ten years:
\begin{itemize}
\item
excess of faint blue galaxies in the field (Tyson 1988) and in cluster
galaxies (Butcher $\&$ Oemler 1978).
\item
post-starburst or E+A galaxies in field and cluster galaxies
(Zabludoff {\it et al.} 1996)
\item
steep rise in the global star formation rate of galaxies from z=0 to 1
(Lilly {\it et al.} 1997)
\end{itemize}

However evidence is growing for the presence of well evolved rich
clusters and cluster galaxies at high redshift (above z=0.5-1;
Dickinson 1997, Mushotzky and Loewenstein 1997), which sets a key
constraint on galaxy evolution models and on structure formation
models as well.

After IRAS, which proved that starbursts emitted mostly in the
infrared (Sanders $\&$ Mirabel 1996), ISOCAM is finding a strong
excess of faint mid-IR (12-18 $\mu m$) emitting galaxies (Cesarsky
1997), demonstrating that dust may affect our view of the star
formation history of the universe. This is confirmed at large
redshifts by the effect of absorption in U and B drop-out selected
galaxies (Pettini {\it et al.}  1997). The fact that galaxy
interaction plays a dominant role both in the hierarchical clustering
scenario and in IRAS ultra-luminous galaxies, strengthens the interest
of studying the effect of the environment on the evolution of cluster
galaxies, whereas the presence of dust requires the use of infrared
telescopes like ISO to complete the analysis.

Observations of galaxy clusters can also set strong constraints on the
evolution of galaxies through the analysis of intra-cluster iron
(Elbaz {\it et al.} 1995), but also on diffuse dust emission outside
galaxies, which could give real-time information on the current status
of the cluster galaxies which would necessarily be replanishing dust
(see below).

\section{Intracluster dust}
Wise {\it et al.} (1993) studied 56 clusters with IRAS and found two
candidate detections of diffuse emission at 60 $\mu m$. Additional
new evidence for the presence of dust outside galaxies has recently
been found by Stickel {\it et al.} (1997), with the detection of an
extended far infrared excess at 120 $\mu m$ at the center of the Coma
cluster. This detection was done using the four pixel camera of
ISOPHOT, C200, on-board ISO, with two 48 arcmin scans rotated about
50$^{\circ}$, each done at 120 $\mu m$ and 185 $\mu m$. 

The authors interpret this excess as thermal emission from
intracluster dust for a total dust mass of 0.6-16
$\times10^{8}~M_{\odot}$. This extragalactic dust must be replenished
continuously since dust is easily destroyed by the hot electrons of
the intracluster medium through sputtering with a timescale of a few
10 million years. It is difficult, however, to see how the dust can be
reinjected at a sufficiently fast rate, although three processes can
be invoked: galactic winds, ram pressure stripping or the recent
accretion of external material by the cluster. In the case of Coma a
poor group of galaxies may have recently merged with the cluster in
its central region.

Because of the possible confusion with the smoothed emission of
individual galaxies, this result would require confirmation with other
clusters. But if confirmed, this result would set strong constraints
on the recent evolution of galaxy clusters.

\section{Mid-Infrared Emission of Galaxies}

The following sections will present results obtained in the
mid-infrared using ISOCAM on-board ISO, in the main broad-band
filters of ISOCAM: LW2 (5-8.5 $\mu m$) centered on 6.75 $\mu m$ and
LW3 (12-18 $\mu m$) centered on 15 $\mu m$. The rest frame mid-IR
emission of galaxies can be divided into three components (Vigroux 1997,
Puget 1997): 

\begin{enumerate}
\item
``carbon composites'': produce the Unidentified Infrared Bands (UIBs) at
6.2, 7.7, 8.6, 11.3 and 12.7 $\mu m$ (the latest was discovered by
ISO) as well as their underlying continuum.
\item 
Warm dust (T$>$150 K): thermal continuum at $\lambda > 10~\mu m$ from
Very Small Grains (VSGs) of dust.
\item
Forbidden lines of ionized gas: NeII (12.8 $\mu m$), NeIII (15.6 $\mu
m$), SIV (10.5 $\mu m$), ArII (7 $\mu m$). 
\end{enumerate}

Only the first component affects the LW2 band, but all three
components emit in the LW3 band, although only HII regions, where the
UV spectrum is hard enough, exhibit strong ionized lines. As the
galaxy redshift increases, LW3 becomes dominated by the first
component while LW2 traces the old stellar component, due to
K-correction.

\section{Star Formation in Nearby Clusters}
Boselli {\it et al.} (1997) have produced a statistical analysis of
the mid-infrared properties of 117 late-type and SO/a galaxies in the
Virgo (99 galaxies detected) and Coma (18 galaxies detected) clusters,
at 6.75 $\mu m$ (87 galaxies detected) and 15 $\mu m$ (72
detections). They used an additionnal sample of ellipticals to
estimate the stellar contribution at 6.75 and 15 $\mu m$ normalized to
the K' flux (consistent with a simple black-body) and substracted it
in the sample of late-types to keep only the dust contribution to the
mid-IR emission. This residual emission originating from dust nicely
correlates with the UV emission of the galaxies at 2000 \AA, which is
a good indicator of star formation. Hence, the rest-frame 6.75 and 15
$\mu m$ emission can be associated with star formation.

\section{Star Formation in z=0.2 Galaxy clusters}
At this redshift, K-correction does not change the physical components
contributing to the LW2 and LW3 ISOCAM filters, which are therefore
equivalent to rest-frame.  A first result of the observation of A1732
(z= 0.193, Pierre {\it et al.} 1997) is that mid-IR emitters avoid the
cluster center, like galaxies providing the blue excess in the
Butcher-Oemler effect, and appear to be morphologically disturbed.

Among 10 galaxies detected at 7 $\mu m$ and probably cluster members,
only 2 were detected at 15 $\mu m$, but with a high 15 over 7 $\mu m$
ratio, hence clearly associated with enhanced star formation. In A1689
(z=0.18), the same excess is found (6 among 9 detections at 15 $\mu m$
show a high 15 over 7 $\mu m$ ratio). More than 20 galaxies were
detected in each cluster at 7 $\mu m$, where ISOCAM is more
sensitive. It is striking that none of the galaxies showing a clear
excess of star formation in the mid-IR can be distinguished from the
rest of the cluster galaxies from its optical colours, e.g. they do
not show any particular blue or red excess. This effect is confirmed
at larger redshift with the galaxies of the Hubble Deep Field detected
by ISO (Aussel 1997), which are spread all over the colour-colour
diagram, confirming that star formation is only partially sampled in
the optical.

\section{Star Formation in z $>$ 0.4 Galaxy clusters}
The K-correction becomes a dominant factor at redshifts above z
$\simeq$ 0.4. The rest-frame emission coming from old stars will be
shifted to the 7 $\mu m$ band, whereas the emission due to UIBs and
their associated continuum, will fall in the 15 $\mu m$ band.

Six clusters were observed at these redshifts ( 3C295, 3C330, 0016+16,
J1888, GHO1322, GHO1603 ) in the frame of a program looking for the
evolution of cluster galaxies as a function of redshift
(P.I. A.Franceschini). Most galaxies were detected at 7 $\mu m$, hence
showing no clear excess of star formation, but the observations are
not complete at faint fluxes (below $\simeq$ 1 mJy). Deeper
integrations have therefore been scheduled.

A search for distant galaxy clusters in the line of sight of bright
quasars (PC1643+4631, Q0000-263) is also in the process of being
reduced (P.I. F.Mirabel).

\section{Conclusions and Perspectives}
ISOCAM deep surveys have been very fruitful, showing a clear excess of
mid-IR emitting galaxies (Deep Survey in the Lockman Hole, Cesarsky
1997, and in the Hubble Deep Field, HDF, Aussel {\it et al.}
1997). These galaxies only show up at faint fluxes (below $\simeq$ 1
mJy) and the observations done at this stage on distant clusters are
too shallow to allow any statistical analysis of such a
population. However, the absence of any correlation between optical
colours and mid-IR fluxes for galaxies in two nearby clusters (A1689
and A1732, at z $\simeq$ 0.2) is interesting since it could show that
ISOCAM is revealing a different population of objects from those
selected in the optical.

The comparison of z $\simeq$ 0.5-1 galaxies in the field and in
clusters will be allowed by the next observations in preparation. The
sample of galaxies detected by ISOCAM within the HDF is already very
promising since they also show no clear optical signature (like a blue
excess) and that they lie above the no-evolution extrapolation from
the IRAS counts.

\end{document}